\pdfoutput=1

\documentclass[11pt]{article}

\usepackage[final]{acl}

\usepackage{times}
\usepackage{latexsym}

\usepackage[T1]{fontenc}

\usepackage[utf8]{inputenc}

\usepackage{microtype}

\usepackage{inconsolata}

\usepackage{inconsolata}
\usepackage{amsmath}
\usepackage{booktabs}
\usepackage{multicol}

\usepackage{multirow}
\usepackage{tcolorbox}
\usepackage{tabularx}
\usepackage{subfigure}
\usepackage{amssymb}
\usepackage{times}
\usepackage{latexsym}

\usepackage{graphicx}
\usepackage{float}
\usepackage{relsize}
\usepackage{algorithm}
\usepackage{algpseudocode}  
\usepackage{xcolor}  

\title{Structure-aware Propagation Generation with Large Language Models for Fake News Detection}

\author{
  \textbf{Mengyang Chen}$^{1,2}$ \quad
  \textbf{Lingwei Wei}$^{1}$\thanks{Corresponding author.} \quad
  \textbf{Wei Zhou}$^{1}$ \quad
  \textbf{Songlin Hu}$^{1,2}$
\\
\\
  $^{1}$Institute of Information Engineering, Chinese Academy of Sciences\\
  $^{2}$School of Cyber Security, University of Chinese Academy of Sciences\\
  \{chenmengyang, weilingwei,  zhouwei, husonglin\}@iie.ac.cn
}
\begin{document}
\maketitle
\begin{abstract}
The spread of fake news on social media poses a serious threat to public trust and societal stability. While propagation-based methods improve fake news detection by modeling how information spreads, they often suffer from incomplete propagation data. Recent work leverages large language models (LLMs) to generate synthetic propagation, but typically overlooks the structural patterns of real-world discussions.
In this paper, we propose a novel structure-aware synthetic propagation enhanced detection (StruSP) framework to fully capture structural dynamics from real propagation. It enables LLMs to generate realistic and structurally consistent propagation for better detection.  StruSP explicitly aligns synthetic propagation with real-world propagation in both semantic and structural dimensions. Besides, we also design a new bidirectional evolutionary propagation (BEP) learning strategy to better align LLMs with structural patterns of propagation in the real world via structure-aware hybrid sampling and masked propagation modeling objective. Experiments on three public datasets demonstrate that StruSP significantly improves fake news detection performance in various practical detection scenarios. Further analysis indicates that BEP enables the LLM to generate more realistic and diverse propagation semantically and structurally.

\end{abstract}
\section{Introduction}

The rapid advancement of online media has led to an alarming rise in fake news, posing significant threats to public trust and societal stability \cite{fisher2016pizzagate,vosoughi2018spread,faris2017partisanship}. 
\begin{figure}[t]
    \centering
    \includegraphics[width=1\linewidth]{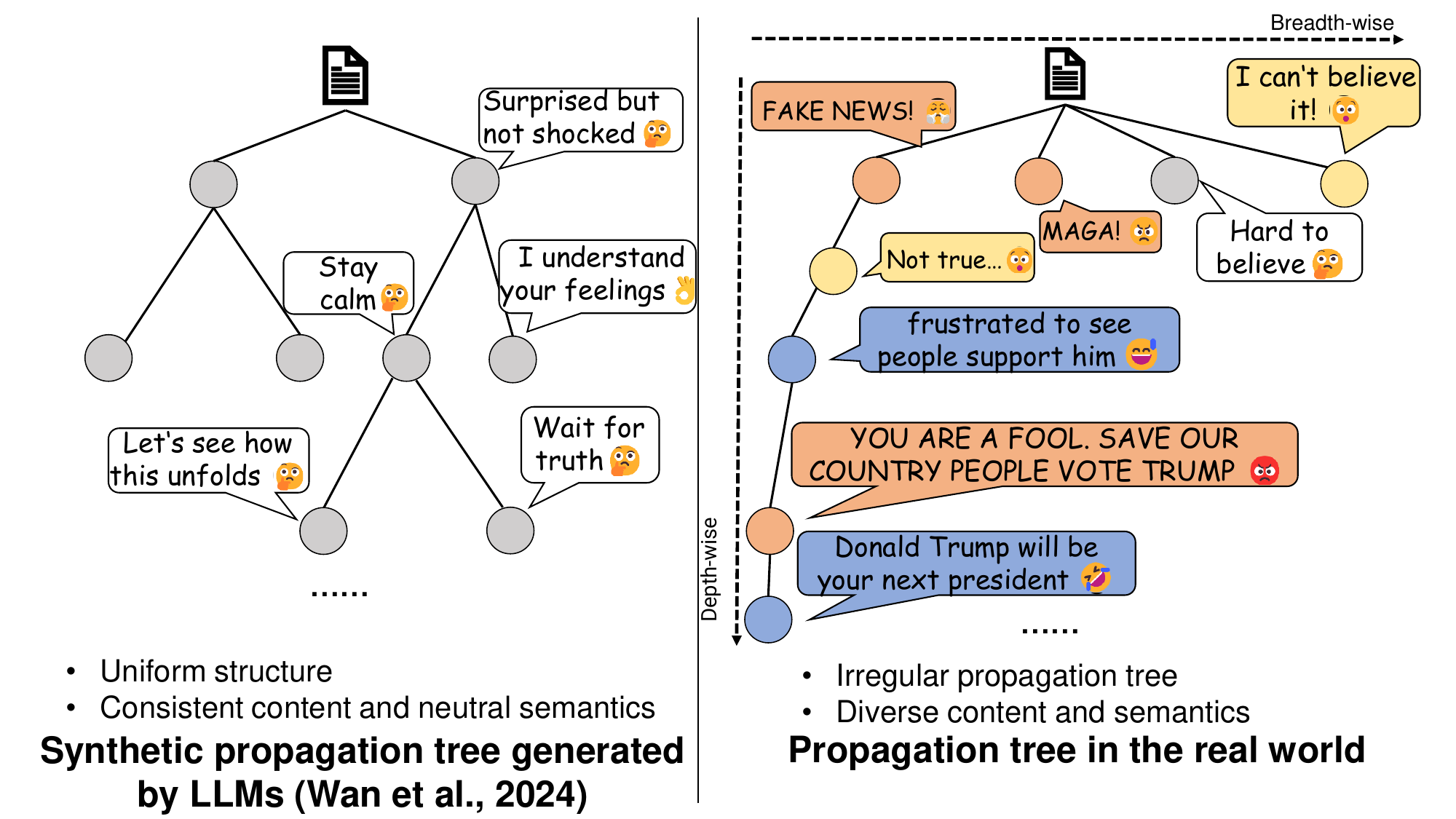}
    \caption{A comparison of LLM-generated propagation (e.g., DELL~\cite{wan2024dell}) and original propagation of the news "\textit{Donald Trump has been disqualified from running for president.}
"}
    \label{fig:intro}
\end{figure}

Existing methods of fake news detection mainly focus on textual content such as news text and contexts   \cite{castillo2011information,ma2015detect,yu2017convolutional}, and propagation information such as interactions between users   \cite{lu2020gcan,su2023mining,Liu2018EarlyDO,bian2020rumor,wei-etal-2021-towards,wu2023decor, chen2024propagation}. 
Despite their promise, propagation-based methods often suffer in incomplete propagation scenarios due to limited data collection and some malicious user interactions on social media \cite{ma2022towards, wei2024transferring}.
Recently, LLMs have shown potential in alleviating data scarcity by simulating user-generated propagation through role-playing approaches \cite{wan2024dell,nan2024let, Liu2024FromST,qiu2025can, yue2024evidence}.

However, these LLM-enhanced methods typically operate only at the semantic level, neglecting the structural patterns of real-world propagation. 
As shown in Figure~\ref{fig:intro}, the generated propagation trees often exhibit overly uniform structures due to the predefined branch probability \cite{wan2024dell} and prompt-induced alignment behavior \cite{denison2024sycophancy, sharmatowards}, failing to capture the irregular branching and hierarchical depth that characterize real-world information spread \cite{zhao2020fake}. In addition, their generated content often lacks emotional diversity and context sensitivity, tending toward overly cautious or generic tones \cite{MuozOrtiz2023ContrastingLP, frisch2024llm}.
The mismatch between the generated and real propagation structures and semantics significantly limits their effectiveness in downstream detection, particularly in early detection and cross-platform generalization. This highlights the need for a structure-aware generation framework that captures both the semantic plausibility and structural dynamics of real-world propagation.

To address the above limitation, we propose a novel \textbf{Stru}cture-aware \textbf{S}ynthetic \textbf{P}ropagation enhanced detection (StruSP) framework to fully capture sufficient features from real-world and LLM-generated synthetic propagation structure. StruSP enriches incomplete propagation trees by generating realistic and structurally consistent propagation paths, thereby improving the effectiveness of propagation-based fake news detection.
To ensure the generated propagation reflects real-world structural dynamics, we introduce a bidirectional evolutionary propagation (BEP) learning strategy to align LLMs with structural patterns of propagation in the real world. BEP consists of two main components. The structure-aware hybrid sampling module first samples propagation substructures via both breadth-wise and depth-wise progression of available propagation trees. Based on these sampled paths, the masked propagation modeling objective captures structural dependencies by reconstructing masked nodes in both forward and backward directions.
This design enables the LLM to effectively learn structural evolution patterns in real-world propagation, equipping it with the ability to enrich incomplete propagation through a structure propagation enhancement module for more accurate detection.

Experiments on three real-world datasets demonstrate that StruSP not only improves detection performance under incomplete propagation conditions but also generates propagation patterns that closely match real data, consistently surpassing baseline methods across structural and semantic metrics.

The contributions of this work can be summarized as follows: 

1) We propose StruSP, a novel structure-aware framework for fake news detection in incomplete propagation scenarios. StruSP enhances fake news detection by generating realistic and structure-aware propagation trees that integrate both semantic and structural signals from partial real propagation.

2) To capture the structural evolution of the real propagation, we introduce a bidirectional evolutionary propagation learning strategy. It enables LLMs to generate structurally diverse and coherent propagation trees.

3) We conduct extensive experiments on three real-world datasets, demonstrating that StruSP significantly improves detection performance and better aligns with real propagation in both structure and semantics.

\section{Related Work}

\paragraph{Fake News Detection}

The goal of detecting fake news is to identify and assess the authenticity of a piece of information. Existing methods for detecting fake news mainly focus on two aspects: textual content and propagation of news.

\textbf{Content-based Fake News Detection Methods} 
extract semantic patterns from news content for detection through feature engineering   \cite{castillo2011information,popat2017assessing,ma2015detect} and a wide array of deep learning architectures, including neural networks   \cite{ruchansky2017csi, karimi2019learning} and pre-trained language models    \cite{kaliyar2021fakebert,jwa2019exbake}.
 Some works also integrate tasks such as stance detection and sentiment analysis with fake news detection, enabling multi-task learning  \cite{luvembe2023dual,hamed2023fake}. Since some fake news creators imitate the style of real news, methods based solely on news content often face limitations. Consequently, some researchers use news comments as a basis for assessing the authenticity of news \cite{shu2019defend, zhang2021mining}.

\textbf{Propagation-based Fake News Detection Methods} capture the propagation patterns of news by modeling the interactions between news and comments into time series   \cite{ma2016detecting,Liu2018EarlyDO} or topological structures such as propagation trees   \cite{ma2018rumor,dou2021rumor} and propagation graphs   \cite{bian2020rumor,wei-etal-2021-towards,wei2022uncertainty}.  Some studies further explore multi-relational interactions between the users and news in the propagation graph   \cite{yuan2020early,dou2021user}.  However, these methods suffer significant performance losses when confronted with scenarios of incomplete propagation \cite{wei2024transferring,ma2022towards}.

\begin{figure*}
    \centering
    \includegraphics[width=1\linewidth]{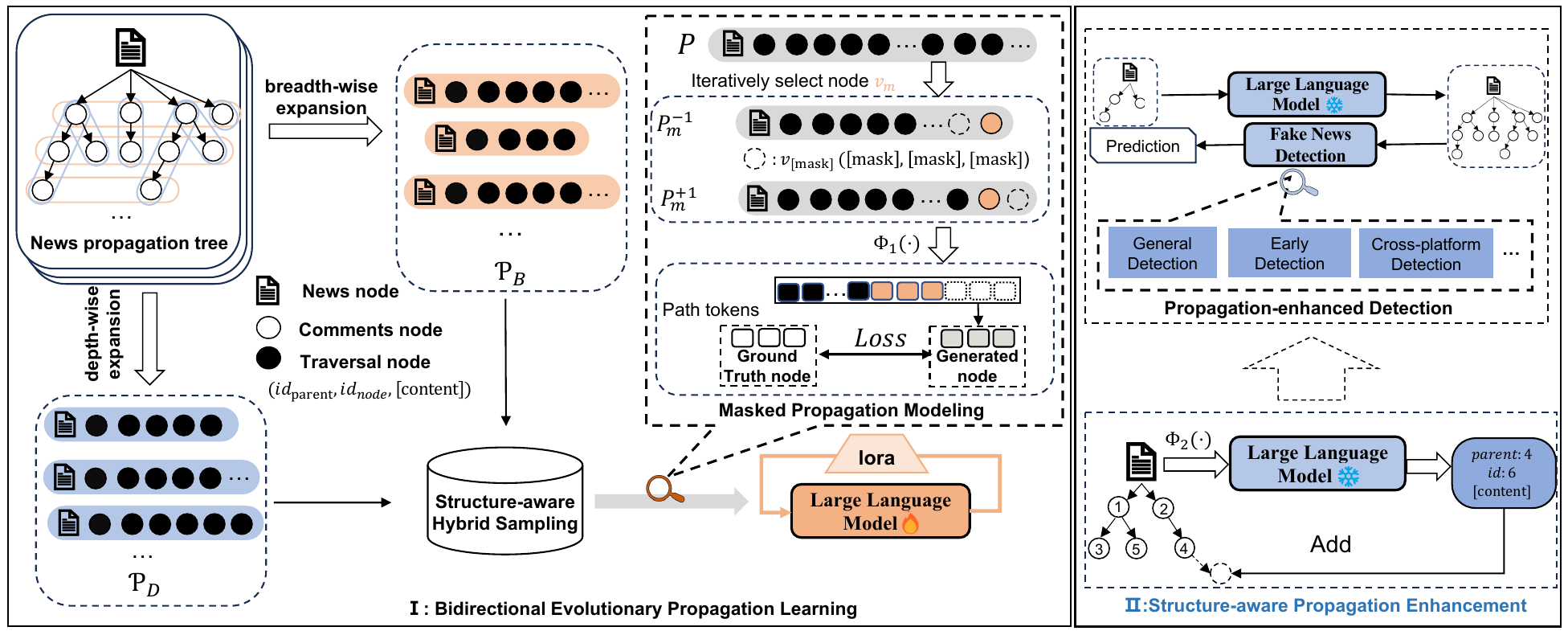}
    \caption{Overall framework of StruSP. We first perform bidirectional evolutionary propagation learning to capture the structural evolution of real propagation. We sample a set of propagation paths for both breadth-wise and depth-wise evolution based on the propagation tree, and then train the LLM to reconstruct masked nodes in forward and backward directions along the sampled path. Then, we generate synthetic propagation to enrich existing propagation for the downstream detection.}

    \label{fig:overview}
\end{figure*}
\paragraph{LLM-based Propagation Generation}
LLMs have been proven to have the potential to simulate human behavior~\cite{argyle2023out} and possess a certain level of social knowledge~\cite{choi2023llms}. 

Recently, some studies have utilized LLMs to simulate social users and generate social contexts~\cite{Gao2023S3SS,Liu2024FromST,Jiang2023SocialLLMMU}. These methods typically leverage LLMs to role-play various users and generate replies to news items, thereby forming synthetic propagation data~\cite{nan2024let,qiu2025can,wan2024dell}.
For instance, \citet{nan2024let} simulated discussions by prompting LLMs to adopt different user identities and respond iteratively. \citet{qiu2025can} further guided the generation process by modeling user behavior through a multilayer perceptron based on historical interactions. 
\citet{wan2024dell} attempted to construct propagation structures by probabilistically controlling whether the LLM comments directly on the news or replies to existing comments. 

However, existing methods either overlook the modeling of propagation structures or generate propagation patterns that do not match real-world structures.
Our proposed StruSP framework explicitly models the structural evolution of propagation and produces propagation trees that better reflect real-world topologies, improving fake news detection performance, especially in scenarios with incomplete propagation.

\section{StruSP Framework}

\paragraph{Problem Statement}
Fake News Detection is to verify the authenticity of a given news article, we take it as a binary classification problem, where each sample is annotated with a ground truth label indicating its authenticity. 
Formally,  Dataset $\mathcal{D}$ consists of $N$ samples and each sample is represented by $\mathcal{G} = (\mathcal{V,E})$, where $\mathcal{V} = \{n,c_1,...,c_N\}$ represents the news $n$ and its comments ($c_1,...,c_N$),  $\mathcal{E}$ represents a set of explicit interactive behaviors (e.g., retweet).
 The task objective of fake news detection is to learn a classifier $f$ to classify samples and determine whether the news is true (labeled as 0) or false (labeled as 1), i.e., 
\begin{equation}
    f\colon\mathcal{G} \longrightarrow y, \quad y \in \{0,1\}.
\end{equation}

\subsection{Overview}
As illustrated in Figure~\ref{fig:overview}, StruSP consists of two key components: bidirectional evolutionary propagation (BEP) learning and structure-aware propagation enhancement (SPE).
The BEP module includes: (1) a structure-aware hybrid sampling strategy that traverses the propagation graph to capture both breadth-wise expansion and depth-wise progression patterns; and (2) a masked propagation modeling objective, which trains the LLM to reconstruct masked nodes along sampled propagation paths, thereby capturing structural dependencies within the diffusion process.
In the SPE module, the trained LLM is used to generate structurally coherent extensions based on incomplete propagation trees.
The synthetic propagation is then integrated into the original propagation structure for enhanced fake news detection.

\subsection{Bidirectional Evolutionary Propagation Learning}

To better model the structural dynamics of news propagation and support realistic propagation generation, we propose a bidirectional evolution propagation learning strategy (BEP), which comprises a structure-aware hybrid sampling mechanism and a direction-aware masked node prediction objective.

\subsubsection{ Structure-aware Hybrid Sampling}

In real-world scenarios, news propagation typically exhibits complex bidirectional dynamics. On one hand, breadth-wise expansion emerges through wide dissemination across social networks (e.g., being shared among diverse user communities), resulting in multi-branched propagation structures. On the other hand, depth-wise progression arises from layered discussions (e.g., multiple rounds of comments and interactions), forming deep sequential chains. Inspired by \citet{tan2023walklm}, we adopt a structure-aware hybrid sampling to model propagation dynamics along both dimensions.
Specifically, given a propagation graph $\mathcal{G}=(\mathcal{V}, \mathcal{E})$, we utilize two graph traversal strategies: Breadth-First Search (BFS) and Depth-First Search (DFS) to encode the bidirectional evolution of the propagation tree:
\begin{equation}
\begin{aligned}
    P^B_\mathcal{G}=BFS(\mathcal{G}), \quad P^D_{\mathcal{G}}=DFS(\mathcal{G}),\\
\end{aligned}
\end{equation}
where $P^B_\mathcal{G}$ captures breadth-wise expansion of $\mathcal{G}$, while $P^D_{\mathcal{G}}$ models depth-wise progression of $\mathcal{G}$. By encoding the propagation tree of each news, we obtain two sets of propagation paths:  breadth-wise expansion paths ($\mathcal{P}_{B}=\{P^B_\mathcal{G}, \mathcal{G} \in \mathcal{D}\}$) and depth-wise progression paths ($\mathcal{P}_{D}=\{P^D_\mathcal{G}, \mathcal{G} \in \mathcal{D}\}$). Each propagation path $P \in \mathcal{P}=\mathcal{P}_{D} \cup \mathcal{P}_{B}$ is represented as a sequence of traversal nodes:
\begin{equation}\label{traversal}
\begin{aligned}
    P&=(v_0,v_1,...,v_{|P| }),\\
v_{i}&=<id_{parent},i,c_i>,
\end{aligned}
\end{equation}
where each traversal node $v_i$ is represented as a triple \footnote{$v_0=(\text{None},0,\text{[news content]})$ denotes the news node in the path.}, $id_{parent} \in \{0,1,...,i-1\}$ represents the parent node index of $v_i$ and $c_i$ indicates the content of $v_i$.

\subsubsection{Masked Propagation Modeling}
While LLMs demonstrate remarkable generalization abilities, domain-specific fine-tuning remains crucial to adapt their broad linguistic knowledge to the unique structural patterns of propagation graphs. 
Drawing inspiration from masked language modeling approaches, we implement a context modeling objective called masked propagation modeling that enhances the LLM's ability to capture the hierarchical evolution patterns inherent in real-world propagation.

Specifically, we iteratively select node $v_m (m\in \{1,2,...,|P|\}$ from $P\in \mathcal{P}$ and mask its preceding and subsequent nodes respectively, generating two sub-paths $P^{-1}_m$ (predict the preceding node $v_{m-1}$) and $P^{+1}m$ (predict the subsequent node $v_{m+1}$) from $P$:
\begin{equation}
\begin{aligned}
P^{-1}_m	
=(v_0,...,v_{m-2},v_{[\text{mask}]},v_{m}),\\
P^{+1}_m=(v_0,...,v_{m-1},v_{m},v_{[\text{mask}]}),
\end{aligned}
\end{equation}
where $v_{[\text{mask}]}$ represents the node to be predicted. These sub-paths $P^{z}_m (z\in \{-1, +1\}$ are then textualized using a predefined prompt template $\Phi_1(\cdot)$ to create training samples:
  \vspace{-2mm}
\begin{tcolorbox}[colback=gray!20, left=1mm, right=1mm, top=1mm, bottom=1mm] 
Given the propagation tree: $P^{z}_m$, please predict the masked comment node (\{'parent node index': '[masked]', 'node index': '[masked]', 'content': '[masked]'\}) in a JSON format as same as other nodes, i.e.,\{parent node index: num, node index: num, content: text\}. 

\end{tcolorbox}
\vspace{-2mm}

The objective essentially constitutes a next-token prediction problem, where the LLM predicts the masked node $v_{[\text{mask}]}$ given input $\Phi_1(P^{z}_m)$ and compares it with the ground truth node $v_{m+z}$. The optimization objective is formulated as:
\begin{equation}
\begin{aligned}
    \mathcal{L}=\sum_{P \in \mathcal{P}}(&\sum_{m=0}^{|P|-1}-\log \mathbb{P}(v_{m+1}|\Phi_1(P^{+1}_m)))\\-&\sum_{m=2}^{|P|}\log \mathbb{P}(v_{m-1}|\Phi_1(P^{-1}_m))).
\end{aligned}
\end{equation}

\subsection{Structure-aware Propagation Enhancement}\label{generation}
Unlike prior approaches \cite{wan2024dell,nan2024let} that simulate propagation solely based on news content, we generate propagation under the guidance of existing propagation using the BEP-trained LLM.
This allows structurally consistent extensions that better reflect real-world propagation patterns and improve downstream detection effectiveness.

Starting with a given news propagation $\mathcal{G}^{'}_0$, we traverse it into a sequence $P_{\mathcal{G}^{'},0}$ like the ones in Equation \ref{traversal} by time order. And then the trained LLM iteratively generates a comment node, which is added to the
propagation sequence:
\begin{equation}\label{enhance}
\begin{aligned}
v^{'}_i&=\text{LLM}(\Phi_2(P_{\mathcal{G}^{'},i-1})),\\
P_{\mathcal{G}^{'},i}&=P_{\mathcal{G}^{'},i-1}\cup \{v^{'}_i\},
\end{aligned}
\end{equation}
where $\text{LLM}$ refers to the BEP-trained LLM, $v^{'}_i$
  represents the generated traversal node as shown in Equation \ref{traversal}. $\Phi_2 (\cdot)$ is the function that encodes
$P^{'}_{n,v_{i-1}}$ into a textual sequence following a predefined prompt template :

  \vspace{-2mm}
\begin{tcolorbox}[colback=gray!20, left=1mm, right=1mm, top=1mm, bottom=1mm] 
Given the propagation tree: $P_{\mathcal{G}^{'},i-1}$, please predict the next comment node in a JSON format as same as other nodes, i.e.,\{parent node index: num, node index: num, content: text\}.

\end{tcolorbox}
\vspace{-2mm}

\subsection{Propagation-enhanced Detection}
  
  Ultimately, we reconstruct the enriched propagation tree 
$\mathcal{G}^{'}_{{k}}$
  by aggregating all node information in $P_{\mathcal{G}^{'},k}$ from Equation \ref{enhance}, where 
$k$
specifies the predefined scale of the number of nodes to generate.

In downstream fake news detection, we use the trained detector $f(\cdot)$ to detect and predict the authenticity label $\hat{y}$ of news using the enriched propagation $\mathcal{G}^{'}_{{k}}$:

\begin{equation}
\begin{aligned}
\hat{y}=f(\mathcal{G}^{'}_{{k}}) .
\end{aligned}
\end{equation}

\section{Experiments Setups}
\begin{table}[t]
    \centering

        \resizebox{0.99\linewidth}{!}{$
    \begin{tabular}{c|rrr}
    \hline 
       Datasets        &  \multicolumn{1}{c}{Twitter}  & \multicolumn{1}{c}{CED} & \multicolumn{1}{c}{PHEME5} \\ 
       \hline

       Number of News      &    1,154 &  3,387&5,801 \\  
 
       Number of True News     &      575	&  1,538&3,829\\ 
     Number of False News     &  579  &1,849&1,973 \\  

       Number of Comments   &59,255   & 1,275,180 &85,408\\ 
       Average number of Comments    &52   & 377&15\\

        \hline 
\end{tabular}
$}
        \caption{The statistics of datasets.     }

    \label{tab:datasets}
\end{table}

\begin{table*}[t]
\centering
\resizebox{0.95\linewidth}{!}{
\begin{tabular}{l|ccccc|ccccc}
\hline 
\multicolumn{1}{c|}{\multirow{2}{*}{Methods}} & \multicolumn{5}{c|}{\multirow{1}{*}{Twitter}} &\multicolumn{5}{c}{PHEME5} \\ 
& Accuracy & Macro-F1  & Recall & Precision & AUC&Accuracy & Macro-F1  & Recall & Precision&AUC\\ \hline 

\multicolumn{1}{l|}{{BERT}} &71.12 &70.86 &71.44 &71.83 &71.13 & 81.83& 79.22&85.56 &87.63 &78.88  \\   
\multicolumn{1}{l|}{{$\text{LLM}_{\text{text}}$}} & 56.00& 54.60& 53.20&55.23 & 57.24& 33.74&28.64&32.78 &34.14 &31.78\\       
\multicolumn{1}{l|}{{dEFEND}} &75.12 &73.56 &75.56 &75.89 &75.13 & 83.24& 82.45&83.54 &92.45 & 89.42 \\ 

\multicolumn{1}{l|}{{$\text{LLM}_{\text{comments}}$}} & 60.75& 60.47& 59.47&65.23 & 61.27& 53.51&52.77 &54.56 &60.27 &54.18 \\ 
\hline
\multicolumn{1}{l|}{{GCN}} & 78.02& 77.67&80.59 &75.39 &86.61 &80.29 &75.16 &81.85 &91.07 & 86.96 \\ 
\multicolumn{1}{l|}{{BiGCN}} &82.76 & 82.71& 84.51& 81.09&90.67 &82.70 &80.28 &86.44 &87.86 &88.44 \\ 
\multicolumn{1}{l|}{{EBGCN}} &83.19 & 82.60& 82.13& 83.44&91.32 &83.89& 80.72& 85.51& 90.71& 89.14\\ 
\multicolumn{1}{l|}{{ RAGCL}} &84.05& 83.72& 83.76& 85.96& 90.88 &84.81& 82.11& 87.29& 89.82&89.25 \\\hline

\multicolumn{1}{l|}{{$\text{LLM}_{\text{propagation}}$}} &48.84& 46.76& 40.24&49.47 & 47.72& 52.14&52.04 &53.94 &54.57 &55.62 \\ 
\multicolumn{1}{l|}{{GenFEND}} & & & & & & & & & & \\ 
\multicolumn{1}{l|}{{\quad $w$/BERT}} & 78.26&75.64 &78.12 &76.64 &74.78 &83.84 &81.23 &87.57 & 89.64& 87.89\\ 
\multicolumn{1}{l|}{{\quad $w$/dEFEND}} & 82.25&82.04 & 83.54 &81.42 &90.70 & 85.07 &84.66 &86.75 &92.46 & 91.63 \\ 

\multicolumn{1}{l|}{{DELL}} & & & & & & & & & & \\ 
\multicolumn{1}{l|}{{\quad $w$/$single$}} &80.17 &79.75 &81.61 &78.91 &88.39 &82.60 &80.75 & 88.84& 84.54&88.89 \\ 
\multicolumn{1}{l|}{{\quad $w$/$vanilla$}} & 80.17&82.55 &83.34 & 76.14& 90.75&82.52 &80.60 &88.50 &84.80 &89.20 \\ 
\multicolumn{1}{l|}{{\quad $w$/$confidence$}} &78.97 &78.10 &83.14 &77.34 &90.21 &83.03 &81.08 &88.62 & 85.60&89.07 \\ 
\multicolumn{1}{l|}{{\quad $w$/$selective$}} &81.22 & 81.02& 81.06&78.96 &86.79 &83.29 & 82.75& 88.80 &89.54 &  89.02\\ \hline
\multicolumn{1}{l|}{{\bf StruSP (Ours)}} & & & & & & & & & & \\ 
\multicolumn{1}{l|}{{\quad $w$/GCN}}& 81.03& 79.75&81.61 & 78.91& 88.39&81.54&80.24 &86.36 &88.41 &80.65 \\ 
\multicolumn{1}{l|}{{\quad $w$/BiGCN}}&84.04& 83.78& \textbf{85.68}& 84.59& 92.08 &84.45& 83.25& 87.85& 88.96&90.70 \\          
\multicolumn{1}{l|}{{\quad $w$/EBGCN}} &84.48 &84.23 & 84.60& 84.85 &\textbf{92.56} &86.21 &84.96 &89.88 &93.12 &90.38\\              
\multicolumn{1}{l|}{{\quad $w$/RAGCL}} &\textbf{85.43} &\textbf{84.84} & 85.05& \textbf{86.22}& 92.45&\textbf{87.76} &\textbf{85.43} &\textbf{90.58} & \textbf{93.17}&\textbf{92.71} \\    \hline 
\end{tabular}
}
\caption{Results (\%) of general fake news detection on Twitter and PHEME5. For each method, we run it five times and report the average results. The results of methods enhanced by StruSP are statistically significant than its baseline model (p-value < 0.05). The best results on each metric are in \textbf{boldface}. }
\label{tab:general}
\end{table*}
\subsection{Datasets}

 We conduct experiments on three public datasets: \textbf{Twitter} \cite{siska-etal-2024-examining}, \textbf{CED} \cite{song2019ced} and \textbf{PHEME5} \cite{zubiaga2016learning}. \textbf{Twitter} contains tweets published on Twitter\footnote{In July 2023, Twitter has been rebranded to X.}, and each tweet is annotated with true of false. \textbf{CED} contains Chinese rumor data scraped from Weibo, including forwarding and comment information related to the original Weibo posts. \textbf{PHEME5} contains collections of rumors and non-rumors released on Twitter during 5 emergency events between 2014 and 2016. 
The statistics of the three datasets are shown in Table \ref{tab:datasets}. Following \citet{chen2025explore}, we divided the datasets into training, validation, and testing sets in a ratio of 7:1:2.

\subsection{Evaluation Metrics}

We evaluate our approach using two categories of metrics. For detection performance, we employ standard classification metrics including Accuracy, Macro-F1, Precision, Recall, and Area Under the ROC Curve (AUC). 

To assess the quality of synthetic propagation, we utilize both structural and semantic metrics. The structural metrics comprise Structural Entropy (SE), Maximum Depth (MD), and Maximum Breadth (MB), which capture the topological characteristics of propagation trees. The semantic metrics include Semantic Consistency (SemC), Sentiment Consistency (SenC), and Semantic Homogeneity (SemH), which measure the coherence of textual content within the propagation. Detailed definitions of these propagation evaluation metrics are provided in Appendix~\ref{appendix:evaluation-metrics}.

\subsection{Baseline}
For the evaluation of our propagation generation methods, we employ the following approaches:

\textbf{BERT} \cite{Devlin2019BERTPO} is a widely used pre-trained language model for fake news detection, with the output from the last layer commonly fed into a classifier. 
\textbf{dEFEND} \cite{shu2019defend} develops a sentence-comment co-attention sub-network for fake news detection.  
\textbf{GCN} \cite{kipf2016semi} applies graph convolutional operations on the news propagation graph to learn news representations.
\textbf{Bi-GCN} \cite{bian2020rumor} models bidirectional propagation graphs based on the news propagation graph for detection.
\textbf{EBGCN} \cite{wei-etal-2021-towards} learns structural features from uncertain propagation using Bayesian graph convolutional networks.
\textbf{RAGCL} \cite{cui2024propagation} 
learns robust rumor representations through adaptive propagation graph contrastive learning.
We utilize the above four propagation-based detection models to evaluate the effectiveness of synthetic propagation for detection.
\textbf{GenFEND} \cite{nan2024let} obtains 30 specific user profiles from three perspectives: gender, age, and education level. Then, LLMs are made to act as these thirty users to comment on news articles.
\textbf{DELL} \cite{wan2024dell} makes LLMs act as designated users to comment on news articles or reply to other comments through an iterative process, thereby generating propagation. 

Additionally, following \citet{chen2025explore}, we evaluate LLMs as fake news detectors, categorizing the models into three types based on input content: LLM$\text{text}$, LLM$\text{comments}$, and LLM$_\text{propagation}$.

\subsection{Implementation Details}
All experiments are conducted on a single NVIDIA A40 GPU with 46GB of memory. The predefined number of generated nodes is set to 30. We implement all baseline methods under the same environment, following the parameter configurations reported in their original papers. Two large language models are used in our study: LLaMa3-8B-Instruct and Qwen3-4B. Unless otherwise specified, we report the results of LLM-based methods using LLaMa3-8B-Instruct.

For training the LLM backbone in StruSP, we construct a joint training set by merging the training portions of the Twitter and PHEME5 datasets. We adopt a parameter-efficient fine-tuning approach using LoRA~\cite{hulora} with a rank of 8, applied to all transformer layers. The model is optimized using the AdamW optimizer with a cosine learning rate schedule, a base learning rate of 5e-5, and a warmup ratio of 0.1. All LLM backbones are trained for 4 epochs with Brain Floating Point 16-bit (BF16) precision enabled.

\section{Experimental Results}

We evaluate the effectiveness of StruSP on three real-world fake news datasets across different detection scenarios (Section \ref {main_results}) and conduct ablation studies to evaluate the effectiveness of each component in StruSP (Section \ref{ablation}).
We further analyze the structural differences between synthetic and real propagation at two levels (Section \ref{Propagation Evaluation}): a macro-level analysis, which compares average metric values across samples, and a micro-level analysis, which examines their distribution at the individual sample level.
We replace the LLM backbone of StruSP to investigate the impact of LLM choice on the framework's performance (Section \ref{Qwen}).

\subsection{Main Results}\label{main_results}
\subsubsection{General Detection}
\begin{figure}[t]
\centering
  \subfigure{\includegraphics[width=0.9\linewidth]{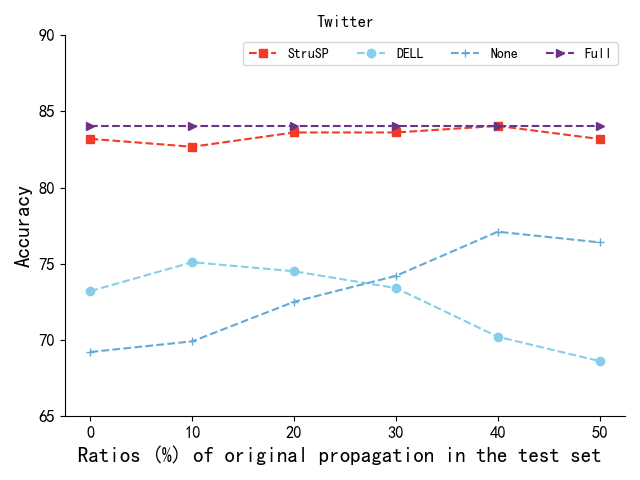}}
  \subfigure{\includegraphics[width=0.9\linewidth]{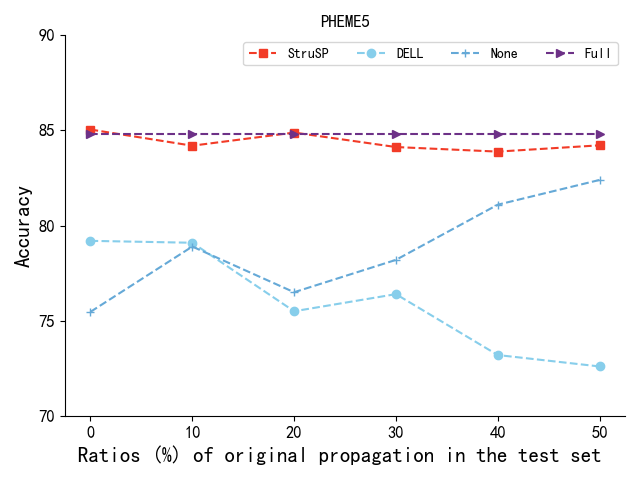}}

\caption{
Results of StruSP and
comparison methods on early detection. RAGCL is the backbone model.
Full refers to using all available propagation data
for prediction, i.e., general detection. None refers to only using limited real-world propagation for early detection. DELL and StruSP use both limited real-world and LLM-generated propagation for early detection.
}
  \label{fig:early}
\end{figure}

Table~\ref{tab:general} shows the performance of baselines and our method in general detection. Our method effectively enhances existing fake news detection methods. Specifically, StruSP $w$/RAGCL achieves the state-of-the-art performance on both datasets, and it gains 2.95\% improvement in accuracy compared to RAGCL on PHEME5.

From the results, we have the following observations. First, compared to GenFEDN and DELL, our propagation-enhanced method performs better, indicating that StruSP generates more informative and structurally realistic propagation trees that empower strong baselines again. Additionally, across all backbone models (GCN, BiGCN, EBGCN, RAGCL), integrating StruSP consistently boosts performance on both Twitter and PHEME5 datasets. This demonstrates that synthetic propagation generated by StruSP effectively complements real propagation in detection.

\subsubsection{Early Detection}

To evaluate the effectiveness of our method in early detection, where only limited propagation data is available, we test RAGCL \cite{cui2024propagation} trained on full propagation data. During testing, only a fixed proportion of the propagation structure is retained to simulate early-stage scenarios \cite{chen2024propagation}. We utilize different methods to enrich the early-stage propagation for testing.

Figure \ref{fig:early} shows early detection results of our StruSP and comparison methods under the RAGCL base model. The results indicate that: 
1) StruSP effectively complements early-stage propagation, achieving detection performance close to the Full setting. This demonstrates its ability to generate structurally and semantically consistent propagation aligned with real-world dynamics.
2) DELL-generated propagation hinders detection performance, highlighting a mismatch between its role-playing generation and actual propagation patterns. This underscores the advantage of structure-aware generation in enhancing early detection.

\subsubsection{Cross-platform Detection}

\begin{table}[t]
    \centering
   
          \resizebox{0.95\linewidth}{!}{

        \begin{tabular}{l|cc|cc}
            \hline 
            \multicolumn{1}{c|}{\multirow{2}{*}{Methods}} & \multicolumn{2}{c|}{\multirow{1}{*}{Twitter $\longrightarrow$ CED}} & \multicolumn{2}{c}{PHEME5 $\longrightarrow$ CED}  \\ 
    
            & Accuracy& Macro-F1 & Accuracy& Macro-F1 \\ \hline 
            \multicolumn{1}{l|}{{RAGCL}} &89.47 & 88.94&73.23 &71.68 \\
            \multicolumn{1}{l|}{{\quad $w$/DELL}} & 88.21&86.34 &78.48&76.92  \\            
            \multicolumn{1}{l|}{\textbf{\quad $w$/StruSP}} & \textbf{92.28}&\textbf{91.56} &\textbf{82.69} &\textbf{80.65}  \\ 
     \hline     
        \end{tabular}
}     \caption{Performance of StruSP and other comparison methods in cross-domain detection. RAGCL is the backbone model.
\textit{Twitter $\longrightarrow$ \text{CED}} refers to training on the source domain (i.e., Twitter) and testing on the target domain (i.e., CED).}    \label{tab:cross}

\end{table}

To investigate generalization ability of StruSP-enhanced fake news detectors in the cross-platform detection, where there is little propagation data on some platform for the platform-specific detectors,  we train RAGCL on two English Twitter datasets separately and test it on the CED dataset from Weibo platform, with the test samples translated into English\footnote{To avoid differences caused by different languages, we translate the text of the samples in the CED test set into English with LLaMa3-8B-Instruct .}. 
 We compare StruSP with DELL by using both to generate propagation data for CED and then perform detection on the generated propagation trees.

As shown in Table~\ref{tab:cross}, RAGCL with StruSP achieves the best performance, demonstrating its ability to retain source-platform (Twitter) propagation dynamics while adapting to target content (Weibo). In contrast, using the original or DELL-generated propagation leads to poor performance, highlighting the difficulty of direct transfer and the importance of structure-aware generation for cross-platform fake news detection.

\subsection{Ablation Study}\label{ablation}
\begin{table}[t]

\centering
        \resizebox{0.95\linewidth}{!}{

        \begin{tabular}{l|cc|cc}
            \hline 
            \multicolumn{1}{c|}{\multirow{2}{*}{Methods}} & \multicolumn{2}{c|}{\multirow{1}{*}{Twitter}} &\multicolumn{2}{c}{PHEME5} \\ 
            & Accuracy & Macro-F1  &Accuracy & Macro-F1  \\ \hline 
   
            \multicolumn{1}{l|}{{\bf StruSP }} &\textbf{85.43}&\textbf{ 84.84}& \textbf{87.76}& \textbf{85.43 }\\   
            \multicolumn{1}{l|}{{\quad $w$/o SHS}} &84.48 &84.25& 85.78& 85.14\\ 
            \multicolumn{1}{l|}{{\quad $w$/o MPM}} &84.56 &84.12& 85.60& 84.02\\  
            \multicolumn{1}{l|}{{\quad $w$/o BEP}} &83.62 &83.22& 83.55& 81.06\\   
            \multicolumn{1}{l|}{{\quad $w$/o SPE}} &84.48 &84.24& 86.12& 85.41\\   
 
            \hline           
 
        \end{tabular}
}

        \caption{Results (\%) comparison between StruSP and its ablative variants. RAGCL is the backbone model. }
    \label{tab:ablation}

\end{table}

To validate the effectiveness of StruSP, we conduct an ablation study of four ablative versions by removing structure-aware hybrid sampling($w$/o SHS), masked propagation modeling ($w$/o MPM), bidirectional evolutionary propagation learning strategy ($w$/o BEP), and structure-aware propagation enhancement ($w$/o SPE). The results are shown in Table \ref{tab:ablation}. It can be observed that the full StruSP $w$/RAGCL achieves better performance on both datasets. The performance drop of $w$/o SHS on both datasets demonstrates the importance of modeling bidirectional evolutionary propagation. Similarly, $w$/o MPM shows degraded results, indicating that providing LLMs with contextual propagation paths aids in capturing propagation dynamics. And the removal of both modules ($w$/o BEP) leads to the most significant decline, confirming that the two components are complementary and jointly crucial for realistic propagation generation. Furthermore, removing the structure-aware propagation enhancement module ($w$/o SPE) results in a noticeable performance drop, indicating that generating propagation solely from news content is less effective. This confirms that leveraging partial real propagation as guidance leads to more informative and structurally aligned synthetic propagation for detection.

\begin{table}[t]
    \centering

        \resizebox{\linewidth}{!}{$
    \begin{tabular}{l|c|c|c|c|c|c}
    \hline 
      \multicolumn{1}{c|}{\multirow{2}{*}{{Methods}}} & 
 \multicolumn{3}{c|}{Structural Metrics} & \multicolumn{3}{c}{Semantic Metrics} 
 \\ 
 &  \multicolumn{1}{c}{SE}  & \multicolumn{1}{c}{MD} & \multicolumn{1}{c|}{MB} &\multicolumn{1}{c}{SemC $\uparrow$ }&\multicolumn{1}{c}{SenC$\uparrow$}&\multicolumn{1}{c}{SemH} \\ \hline

       Orginal      & 0.94    & 3.56 &14.17 & - & -&  0.85\\        
       GenFEND    &   -  & - &- & 0.89 & 0.53&0.91\\        
     DELL    &  1.75  & 4.71   & 10.09 & 0.91 & 0.50& 0.92 \\        
       {{LLM}}     & 1.54    & 4.10 &11.13 & 0.94 & 0.79& 0.88 \\       

       StruSP   &  0.84   & 3.99 &13.72 &0.97  &0.85& 0.86  \\

        \hline 
\end{tabular}
$}
        \caption{Results of macro-level propagation analysis on the combined Twitter and PHEME5 datasets. Structural Entropy (SE), Maximum Depth (MD), and Maximum Breadth (MB) evaluate the structural features of propagation. Semantic Consistency (SemC), Sentiment Consistency (SenC), and Semantic Homogeneity (SemH) evaluate the semantic features of propagation. }

    \label{tab:macro}
\end{table}
\begin{figure}[t]
\centering
  \subfigure[Structure Entropy]{\includegraphics[width=0.45\linewidth]{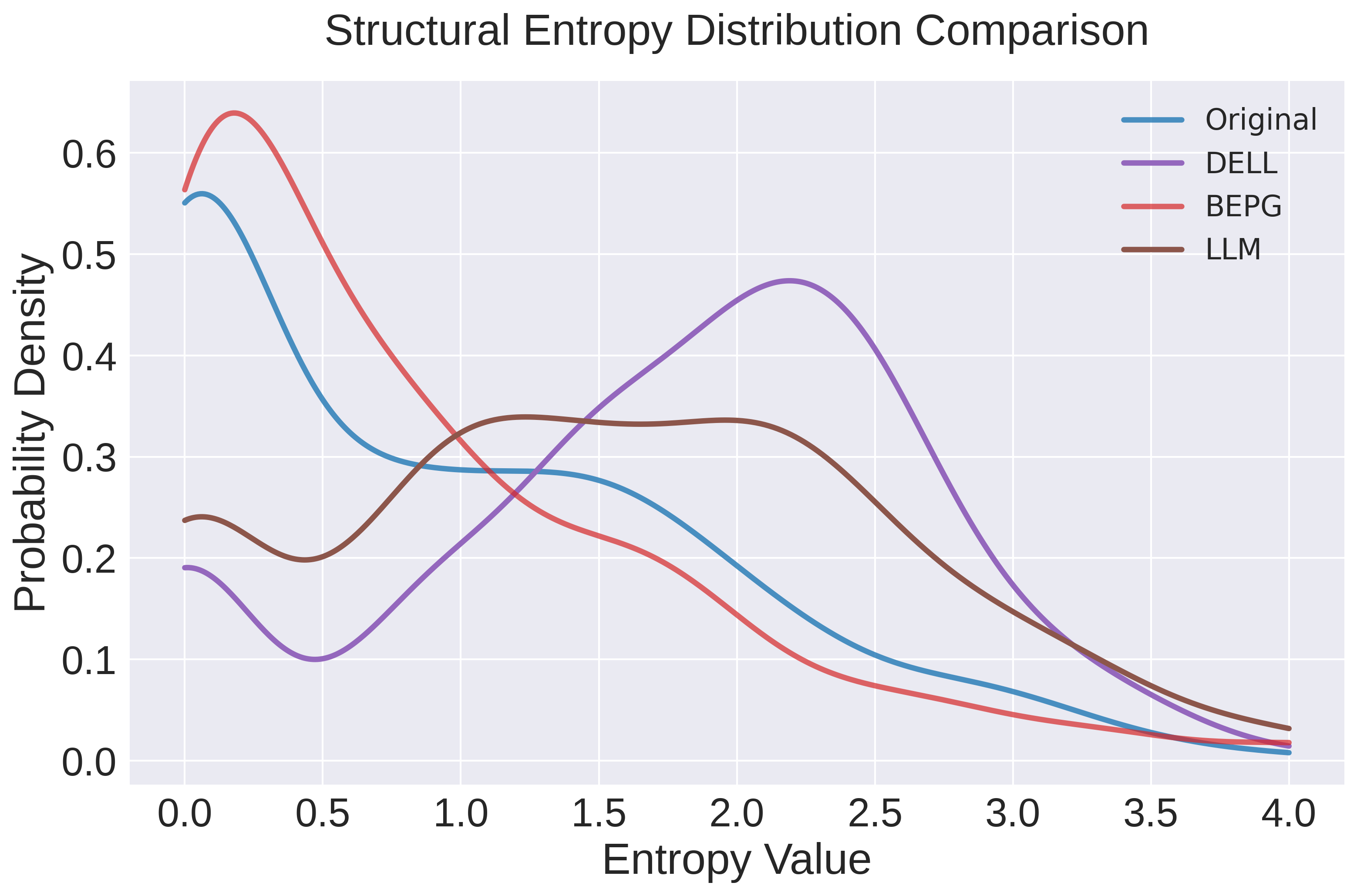}}
  \subfigure[Max Depth]{\includegraphics[width=0.45\linewidth]{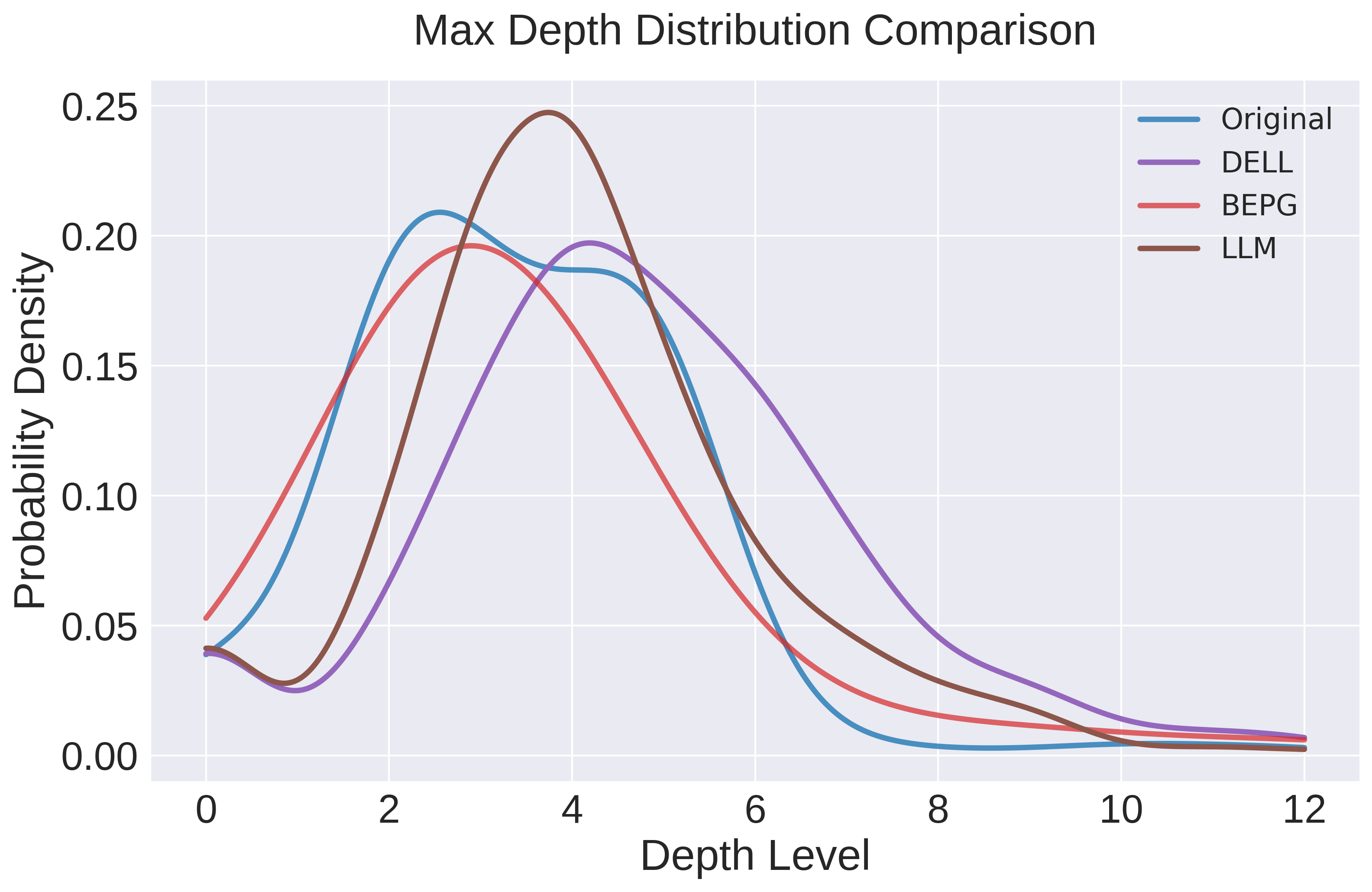}}
  \subfigure[Max Breadth]{\includegraphics[width=0.45\linewidth]{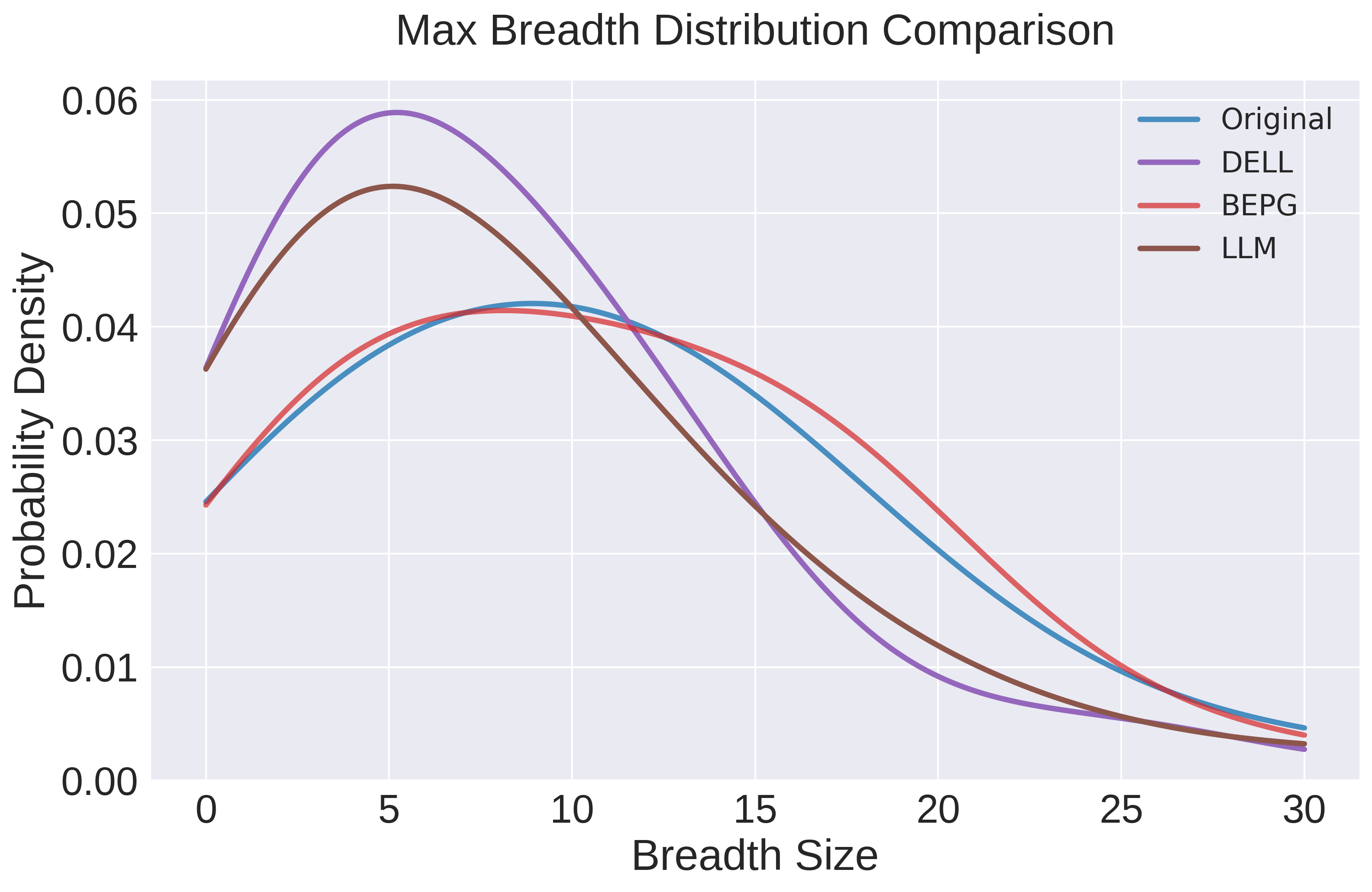}}
  \subfigure[Semantic Homogeneity]{\includegraphics[width=0.45\linewidth]{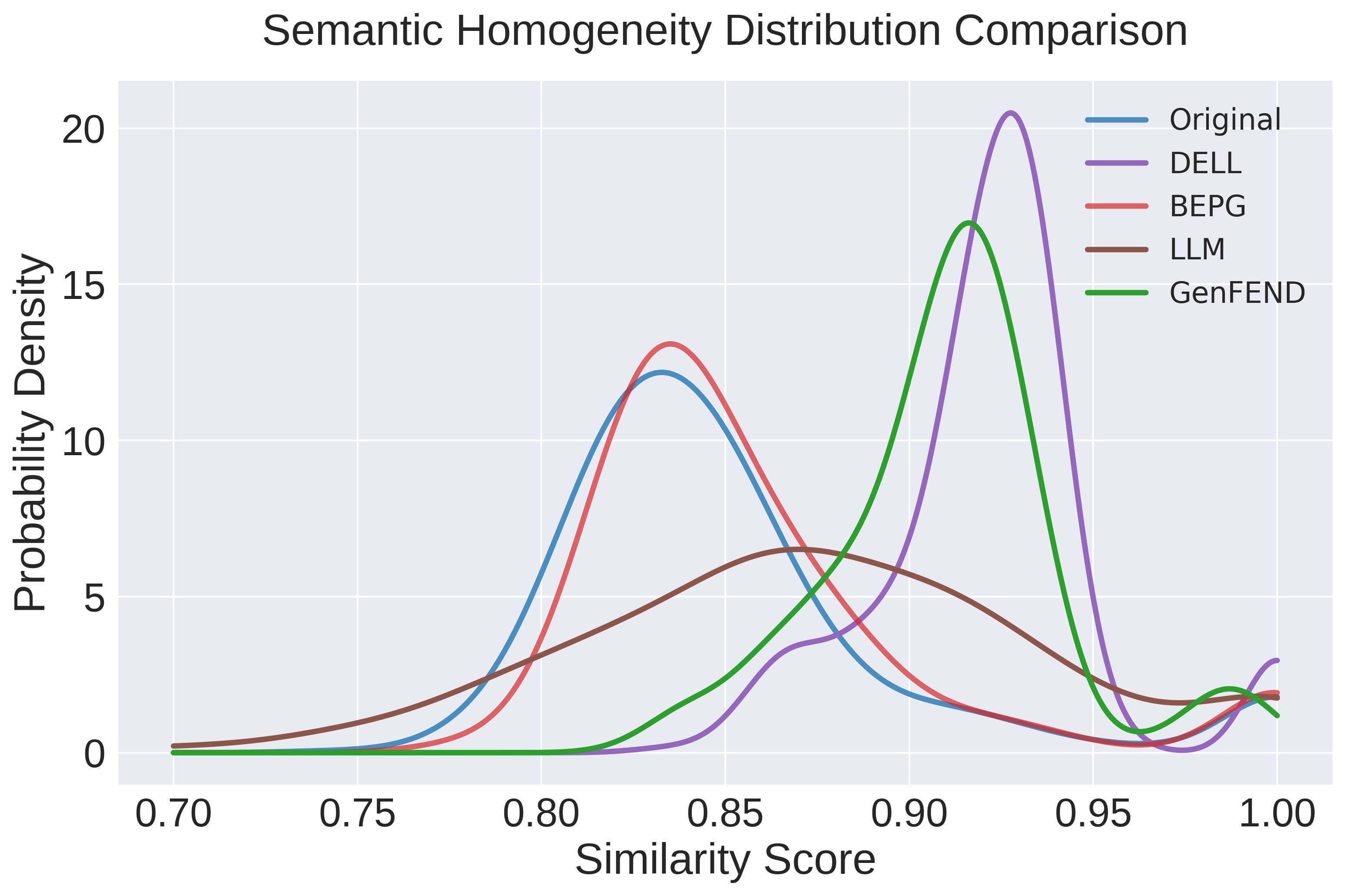}}

\caption{
Results of micro-level propagation analysis on the combined Twitter and PHEME5 datasets. The distributions of Structural Entropy, Instance-level Depth/Breadth, and Semantic Homogeneity across propagation generated by different methods
}
  \label{fig:micro}
\end{figure}

\subsection{Propagation Evaluation}\label{Propagation Evaluation}

We evaluate the quality of propagation generated by StruSP and other comparison methods. \textbf{LLM} refers to the use of an unfine-tuned LLaMa3-8B-Instruct model to generate propagation, following the approach described in Section~\ref{generation}. We conduct both macro-level and micro-level analyses. The macro-level evaluation assesses the overall structural and semantic similarity to real-world propagation, while the micro-level evaluation focuses on intra-propagation variation. Detailed definitions of all evaluation metrics are provided in Appendix~\ref{appendix:evaluation-metrics}.

The evaluation results are shown in Table \ref {tab:macro} and Figure \ref{fig:micro}. It can be observed that synthetic propagation generated by StruSP best aligns with real-world propagation, outperforming non-fine-tuned LLMs and role-playing methods (e.g., GenFEND, DELL) in both structural and semantic metrics. It indicates that the effectiveness of StruSP in guiding LLMs to generate realistic and informative propagation. Moreover, StruSP shows higher similarity to real propagation across semantic metrics compared to non-fine-tuned baselines, confirming the benefit of incorporating real-world propagation signals during training.

\subsection{Performance with Different LLM Backbones}\label{Qwen}

\begin{table}[t]
\centering
\small
\setlength{\tabcolsep}{4pt} 
\renewcommand{\arraystretch}{1.15} 
        \resizebox{0.85\linewidth}{!}{

\begin{tabular}{l|cccc}
\hline
\multirow{2}{*}{Method} & \multicolumn{2}{c}{Twitter} & \multicolumn{2}{c}{PHEME5} \\
& Accuracy & Macro-F1 & Accuracy & Macro-F1 \\
\hline
\multicolumn{5}{l}{\textit{LLaMa3-8B as backbone}} \\
GCN & 81.03 & 79.75 & 81.54 & 80.24 \\
BiGCN & 84.04 & 83.78 & 84.45 & 83.25 \\
EBGCN & 84.48 & 84.23 & 86.21 & 84.96 \\
RAGCL & \textbf{85.43} & \textbf{84.84} & \textbf{87.76} & \textbf{85.43} \\
\hline
\multicolumn{5}{l}{\textit{Qwen3-4B as backbone}} \\
GCN & 80.72 & 78.14 & 81.82 & 80.56 \\
BiGCN & 83.62 & 83.19 & 84.07 & 83.16 \\
EBGCN & 84.48 & 84.23 & 85.96 & 84.87 \\
RAGCL & \textbf{84.91} & \textbf{84.42} & \textbf{87.07} & \textbf{86.48} \\
\hline
\end{tabular}
}
\vspace{2mm}
\caption{Results (\%) of StruSP with different LLM backbones and GNN variants in general fake news detection. All models are evaluated on Twitter and PHEME5 datasets.}
\label{tab:LLM}
\end{table}

To validate the generalization of our method for different LLMs, we compare the performance of StruSP with LLaMa3-8B-Instruct and Qwen3-4B as the backbone and conduct a comprehensive performance comparison.  As shown in Table \ref{tab:LLM}, LLaMa3-8B-Instruct and Qwen3-4B achieved comparable performance. It indicates that StruSP is robust to the choice of backbone LLM, as both LLaMa3-8B-Instruct and Qwen3-4B can effectively generate synthetic propagation data to enhance fake news detection. Moreover, our method shows the best detection performance with the different LLM backbones.

\section{Conclusion}
This paper proposes a structure-aware synthetic propagation enhanced fake news detection framework (StruSP). By employing a bidirectional evolutionary propagation learning strategy, StruSP enables LLMs to generate realistic and informative propagation trees and enrich the existing incomplete propagation tree. Experiments on three datasets demonstrate that StruSP significantly improves fake news detection performance in different incomplete propagation settings and produces realistic and diverse propagation. 

\section{Acknowledgements}
This work was supported by the National Natural Science Foundation of China (No.U24A20335), the China Postdoctoral Science Foundation (No.2024M753481), and Youth
Innovation Promotion Association CAS. The authors thank the anonymous reviewers and the metareviewer for their helpful comments.

\section*{Limitations}
As an initial attempt to generate synthetic propagation that structurally aligns with real-world diffusion patterns, the proposed StruSP framework presents several limitations. First, it relies on observed propagation data to train the LLM-based generator, which makes its performance sensitive to the quality and coverage of the training data. Second, the generation quality is closely tied to prompt design; less expressive or overly generic prompts may struggle to capture complex structural dependencies. Lastly, StruSP does not explicitly model user-level behaviors or social dynamics, potentially limiting the realism and personalization of the generated content, particularly in emotionally charged or user-driven conversations.

\section*{Ethics Statement }

Our proposed method, StruSP, leverages large language models (LLMs) to generate synthetic propagation structures for the purpose of enhancing fake news detection under incomplete propagation scenarios. All generated content is used solely for research purposes and is not intended for public dissemination.

While our approach improves detection performance, we acknowledge the potential misuse of synthetic propagation generation for malicious purposes, such as falsifying social media diffusion. To mitigate this, we emphasize that our system is designed for controlled research and evaluation within the context of fake news detection.

We also recognize that LLM-generated content may reflect unintended biases or sentiments. To reduce this risk, we employ structure-aware training grounded in real-world data and evaluate outputs using semantic and sentiment alignment metrics.

Overall, this study aims to advance the understanding and mitigation of fake news on social platforms, and we encourage the responsible use of the techniques proposed.


\clearpage
\appendix

\section{Propagation Evaluation Metrics}\label{appendix:evaluation-metrics}

We provide detailed definitions and calculation methods for the metrics used in evaluations of propagation quality.

\textbf{Structural Entropy (SE)} 
quantifies uncertainty in node degree distribution using Shannon entropy:
\begin{equation}
\text{SE} = - \sum_{k} p_k \log p_k,
\end{equation}
where \( p_k \) is the proportion of nodes with degree \( k \).

\vspace{0.5em}
\textbf{Max Depth (MD)} measures the maximum depth of a single propagation instance.   \\

\textbf{Max Breadth (MB)} measures the maximum number of nodes at any depth level.   \\

\textbf{Semantic Consistency (SemC)} measures the average semantic alignment between the generated propagation tree $\mathcal{G}^{'}$ and the corresponding original propagation tree $\mathcal{G}$ by comparing their aggregated semantic representations:
\begin{equation*}
\small
\begin{split}
\text{SemC} = \frac{1}{N} \sum_{i=1}^{N} 
\cos\Bigg(&
\frac{1}{|V^{(\mathcal{G}_i^{'})}|} \sum_{v \in V^{(\mathcal{G}_i^{'})}} \text{emb}(v), \\
& \frac{1}{|V^{(\mathcal{G}_i)}|} \sum_{v \in V^{(\mathcal{G}_i)}} \text{emb}(v)
\Bigg),
\end{split}
\end{equation*}
where $N$ denotes the number of samples. $V_i^{(\mathcal{G})}$ and $V_i^{(\mathcal{G}^{'})}$ are the node sets of $\mathcal{G}_i$ and $\mathcal{G}_i^{'}$ respectively. $\text{emb}(v)$ is the BERT-based embedding of node $V$. The cosine similarity is computed between the average embedding of each tree pair.

\vspace{0.5em}
\textbf{Sentiment Consistency (SenC)} measures the alignment of overall sentiment across generated and original propagation. It reflects whether the generated propagation $\mathcal{G}^{'}$ preserves the dominant sentiment polarity of the original ones $\mathcal{G}$. 
\begin{equation}
\small
\text{SenC} =
\frac{1}{N} \sum_{i=1}^{N} \mathbb{I}\left( \text{MajSent}(\mathcal{G}^{'}_i) = \text{MajSent}(\mathcal{G}_i) \right),
\end{equation}
where $N$ denotes the number of samples. $\text{MajSent}(\mathcal{G}_i)$ represents the majority sentiment label (e.g., Positive or Negative) of the 
$i$-th tree in the datasets. The function $\mathbb{I}(\cdot)$ is the indicator function that returns 1 if the two labels are equal, and 0 otherwise. Sentiment labels are obtained using a pretrained sentiment classifier\footnote{https://huggingface.co/distilbert/distilbert-base-uncased-finetuned-sst-2-english} applied to each comment in the propagation tree. The majority sentiment of a tree is determined by the most frequent label among its nodes.

\vspace{0.5em}
\textbf{Semantic Homogeneity (SemH)} measures pairwise coherence among all comments in a propagation, the  cosine similarity is used to calculate the semantic homogeneity:
\begin{equation}
\small
\text{SemH} =\\
\frac{2}{|V|(|V|-1)} \sum_{i<j} \cos\left(\text{emb}(v_i), \text{emb}(v_j)\right)
\end{equation}

\section{Prompt Design and Robustness Analysis}
\label{sec:prompt_robustness}

\subsection{Robustness Through Fine-tuning and Structure}
\label{sec:robustness_mechanism}

Our framework achieves robustness against prompt variations through two core design principles:

\textbf{1. Deep Structural Learning via BEP:} The Bidirectional Evolutionary Propagation learning strategy (Section 3.2 in the main paper) explicitly trains the LLM to become an expert in propagation dynamics. This process deeply ingrains structural awareness into the model's parameters, making its behavior inherently stable and far less sensitive to minor prompt variations compared to methods relying solely on in-context learning.

\textbf{2. Formal API Design:} By treating prompts as a formal API with structured JSON I/O (detailed in Section~\ref{sec:json_format}), we ensure that the LLM engages with the task's logic directly. This structured interface minimizes ambiguity and ensures consistent interpretation regardless of natural language variations in the prompt.

These design choices collectively ensure that our method's performance is robust and reliable across different prompt formulations.

\subsection{Structured JSON I/O Format}
\label{sec:json_format}

Our framework employs a structured JSON format for both input and output:

\vspace{-2mm}
\begin{tcolorbox}[colback=gray!20, left=1mm, right=1mm, top=1mm, bottom=1mm]
\textbf{Input:} \\
Propagation tree with nodes: [\{\texttt{node\_index}: 0, \texttt{parent\_index}: -1, \texttt{content}: "text"\}, ...]

\textbf{Output:} \\
Predicted node: \{\texttt{parent\_node\_index}: num, \texttt{node\_index}: num, \texttt{content}: "text"\}
\end{tcolorbox}
\vspace{-2mm}

This structured approach transforms the interaction into a clear, machine-readable function call, ensuring reliable and predictable outputs while minimizing ambiguity in the LLM's interpretation of the task.

\subsection{Prompt Sensitivity Analysis}
\label{sec:prompt_sensitivity}

To demonstrate the robustness of our Bidirectional Evolutionary Propagation (BEP) learning strategy against prompt variations, we conducted comprehensive experiments using three different prompt formulations on the Twitter dataset.

\subsubsection{Prompt Variants}
We evaluated the following prompt variants:

\textbf{P1 (Structured - Ours):} \textit{"Given the propagation tree: \{tree\}, please predict the next comment node in a JSON format as same as other nodes, i.e., \{parent node index: num, node index: num, content: text\}."}

\textbf{P2 (Minimal):} \textit{"Given the propagation tree: \{tree\}, please predict the next comment node."}

\textbf{P3 (Detailed):} \textit{"Given the propagation tree: \{tree\}, please carefully analyze the structural patterns and semantic context, then predict the next comment node that maintains both structural consistency and semantic coherence in a JSON format, i.e., \{parent node index: num, node index: num, content: text\}."}

\subsubsection{Experimental Results}

Table~\ref{tab:prompt_robustness} presents the performance comparison across different prompt variants, demonstrating the stability of our approach.

\begin{table}[t]
\centering
\resizebox{1\linewidth}{!}{
\begin{tabular}{ccccc}
\toprule
Method & Accuracy $\uparrow$ & Macro-F1 $\uparrow$ & $|\text{SE}_{\text{ori.}}-\text{SE}_{\text{syn.}}|$ $\downarrow$& SemC $\uparrow$ \\
\midrule
StruSP w/P1 (Ours) & 85.43 & 84.84 & 0.10 & 0.97 \\
StruSP w/P2 & 84.91 & 84.45 & 0.14 & 0.97 \\
StruSP w/P3 & 85.12 & 84.36 & 0.10 & 0.97 \\
\midrule
DELL & 81.22 & 81.02 & 0.81 & 0.91 \\
GenFEND & 82.25 & 82.04 & - & 0.89 \\
\bottomrule
\end{tabular}
}
\caption{Performance comparison across different prompt variants on the Twitter dataset. $|\text{SE}_{\text{ori.}}-\text{SE}_{\text{syn.}}|$ denotes the absolute value of the difference between the structural entropy score of the original propagation ($\text{SE}_{\text{ori.}}$) and the structural entropy score of the synthetic propagation ($\text{SE}_{\text{syn.}}$). The results demonstrate that our BEP learning strategy ensures robust performance regardless of prompt formulation.}
\label{tab:prompt_robustness}
\end{table}

\subsubsection{Analysis and Discussion}

The experimental results reveal several key insights:

\begin{enumerate}
    \item \textbf{Minimal Performance Variation:} The performance difference between prompt variants is marginal (less than 0.52\% in ACC and 0.39\% in Macro-F1), demonstrating that our BEP learning strategy successfully reduces sensitivity to prompt formulation.
    
    \item \textbf{Consistent Structural Understanding:} All prompt variants maintain similar Structural Entropy and identical Semantic Consistency with original propagation, indicating that the model's understanding of propagation dynamics is deeply ingrained through fine-tuning rather than dependent on prompt engineering.
    
    \item \textbf{Superiority over Baselines:} Even with the minimal prompt (P2), our method outperforms baseline approaches by significant margins, confirming that the robustness stems from our fine-tuning strategy rather than prompt sophistication.
\end{enumerate}

\section{Computational Cost Analysis}
\label{sec:computational_cost}

We employed Parameter-Efficient Fine-Tuning (PEFT) using LoRA (rank=8) to significantly reduce the training cost while maintaining model performance. Fine-tuning the LLaMa3-8B-Instruct model on the combined Twitter and PHEME5 datasets took approximately 4 hours on a single NVIDIA A40 GPU, representing a one-time, manageable cost. This approach reduces the number of trainable parameters to approximately 0.1\% of the original model size, making the fine-tuning process highly memory-efficient and accessible even in resource-constrained environments.

During inference, generating 30 synthetic comments for a single news propagation tree takes on average 1 minute on a single NVIDIA A40 GPU. This modest one-time training cost creates a powerful, reusable generator that can produce unlimited amounts of high-quality training data, enabling significant performance gains for lightweight GNN detectors in low-resource scenarios. For example, augmenting 1000 propagation trees would require approximately 16.7 GPU hours but would yield 30,000 high-quality training samples, demonstrating an excellent cost-benefit ratio for practical deployment.

\section{Generation Process and Automated Quality Assurance}
\label{sec:generation_details}

Our generation process employs carefully selected parameters to balance diversity with coherence. We generate 30 nodes per propagation tree using the fine-tuned LLaMA3-8B model with LoRA adaptation. For decoding, we use nucleus sampling (top-p = 0.9) with a temperature of 0.6, allowing up to 3 retry attempts per node generation. These parameters were empirically determined to ensure synthetic propagation patterns remain realistic while providing sufficient variability for effective data augmentation.

The quality of generated data is ensured through a rigorous automated validation pipeline, as detailed in Algorithm~\ref{alg:validation}. Each generated node must pass through three sequential validation gates before being accepted into the propagation tree. The \textbf{Syntactic Filter} ensures the LLM's output is well-formed JSON, triggering re-generation for any malformed responses. The \textbf{Structural Filter} validates topological integrity by checking for valid parent references, preventing self-loops, ensuring unique node identifiers, and maintaining proper tree structure without cycles. Finally, the \textbf{Content Filter} ensures semantic quality by filtering out empty content, boilerplate refusal messages, repetitive text, and responses shorter than a minimum threshold.

\begin{algorithm}[htbp]
\caption{Automated Node Validation Pipeline}
\label{alg:validation}
\begin{algorithmic}
\Require Existing propagation tree $P_{current}$, $\text{LLM}_{generator}$, $\text{Max\_retries} = 3$
\Ensure A new valid node $v_{new}$ or Failure
\Function{GenerateValidNode}{$P_{current}$}
    \For{$i = 1$ to $\text{Max\_retries} + 1$}
        \State \textcolor{gray}{\textit{//  Generate a candidate node}}
        \State $\text{raw\_output} \gets \text{LLM}_{generator}(P_{current})$
        
        \State \textcolor{gray}{\textit{// --- Validation Gate 1: Syntactic Filter ---}}
        
        \State \textbf{try:}
        \State \quad $\text{node\_json} \gets \text{ParseJSON}(\text{raw\_output})$
        \State \textbf{except} JSONDecodeError:
        \State \quad \textbf{continue} \Comment{Retry if output is not valid JSON}
        
        \State \textcolor{gray}{\textit{// --- Validation Gate 2: Structural Filter ---}}
        
        \If{\textbf{not} IsStructurallyValid($\text{node\_json}, P_{current}$)}
            \State \textbf{continue} \Comment{Retry if parent\_id or node\_id is invalid}
        \EndIf
        
        \State \textcolor{gray}{\textit{// --- Validation Gate 3: Content Filter ---}}

        \If{IsContentInvalid($\text{node\_json}[\text{'content'}]$)}
            \State \textbf{continue} \Comment{Retry if content is empty, refusal, etc.}
        \EndIf
        
        \State \textcolor{gray}{\textit{// --- Success: Node is valid ---}}
        
        \State \textbf{return} CreateNode($\text{node\_json}$)
    \EndFor
    
    \State \textcolor{gray}{\textit{// If all retries fail, return Failure}}
    
    \State \textbf{return} Failure
\EndFunction
\end{algorithmic}
\end{algorithm}



\end{document}